\documentstyle[12pt,aaspp4]{article}

\def\simlt{\lower.5ex\hbox{$\, \buildrel < \over \sim \,$}}
\def\simgt{\lower.5ex\hbox{$\, \buildrel > \over \sim \,$}}
\def\sun{\mbox{$_\odot$}}
\def\etal{{\it et al. }}

\begin{document}

\title{The Mount Stromlo Abell Cluster Supernova Search}

\author{David J. Reiss}
\affil{Dept. of Astronomy, University of Washington \\
email: reiss@astro.washington.edu}
\authoraddr{Phys. Astron. Bldg. Box 351580; Seattle, WA; 98195}

\author{Lisa M. Germany}
\affil{Mount Stromlo and Siding Spring Observatories \\
email: lisa@mso.anu.edu.au}
\authoraddr{Private Bag; Weston Cr. P.O.; ACT 2611; Australia}

\author{Brian P. Schmidt}
\affil{Mount Stromlo and Siding Spring Observatories \\
email: brian@mso.anu.edu.au}
\authoraddr{Private Bag; Weston Cr. P.O.; ACT 2611; Australia}

\and

\author{C. W. Stubbs}
\affil{Depts. of Physics and Astronomy, University of Washington, \\
Mount Stromlo and Siding Spring Observatories \\
email: stubbs@astro.washington.edu}
\authoraddr{Phys. Astron. Bldg. Box 351580; Seattle, WA; 98195}

\begin{center}
http://msowww.anu.edu.au/\~{ }reiss/Abell\_SNSearch
\end{center}


\begin{abstract}
  We have initiated a three-year project to find supernovae (SNe) in a
  well-defined sample of high-density southern Abell clusters with
  redshifts $z\leq0.08$.  These observations will provide a
  volume-limited sample of SNe~Ia to more than a magnitude below their
  peak brightness, and will enable us to: (1) measure the luminosity
  function of SNe, (2) further explore the correlation of light curve
  shape with the absolute luminosity of SNe~Ia to better understand
  SNe~Ia as distance indicators, (3) measure SN rates, (4) measure the
  bulk motion of the Local Group using SNe~Ia, and (5) directly
  compare SN~Ia distances to brightest cluster galaxy distances.  We
  use the MaCHO wide-field 2-color imager on the 1.3m telescope at
  Mount Stromlo to routinely monitor $\sim 12$ clusters per week.  We
  describe our technique for target selection and scheduling search
  observations, and for finding and identifying SN candidates. We also
  describe the results from the first year of our program, including
  the detection of 19 SNe, several RR-Lyrae variables, and hundreds of
  asteroids.
\end{abstract}

\section{Introduction and Motivation}
\label{sec:intro}
As extremely luminous point sources which represent a discrete
physical event, SNe are attractive indicators of extragalactic
distances.  The CTIO/Cal\'an group (\cite{hetal95}; \cite{hetal96a})
demonstrated that Type~Ia SNe are not standard candles but instead
show a range in peak luminosity of approximately one magnitude in V.
However, they also proved that there exists a tight correlation
between decline rate and peak brightness for SNe~Ia in the sense that
more luminous events decline more slowly than intrinsically faint ones
(\cite{phil93}).  This property of SN~Ia light curves provides a means
to accurately estimate the intrinsic luminosity of the SN, thereby
sharpening the precision of SNe~Ia as standard candles to better than
$\sigma_V<0.2^m$.  A somewhat more elaborate method developed by
\cite{rpk95b}, 1996\nocite{rpk96} uses multi-band light curve shapes
to estimate the luminosity of type~Ia SNe and also predicts distance
uncertainties and extinction for individual SNe.  Using a sample of 20
SNe drawn from \cite{hetal96b} and their own observations, they were
able to improve the distance estimate for each, decreasing the scatter
in the Hubble diagram to $\sigma=0.12^m$ (distance error of 6\%).

For type~Ia SNe to be useful for measuring distances, they must be
discovered near peak brightness.  Searches based on infrequent
observations typically produce SNe that are too old to obtain good
followup photometry and measure well-sampled light curves.  In a
systematic search in which each field is revisited periodically, the
ages of any discovered SNe are constrained to be less than the time
since the field was last observed.  Such frequent visits enable more
accurate light-curve shape measurements, and thus, more accurate
distances.  Systematic searches for distant SNe ($z\simgt 0.4$) are
currently being carried out by \cite{perletal97} and \cite{highz97}
while nearby searches are being conducted by groups at the Beijing
Astronomical Observatory (\cite{baoref}), the Perth Observatory
(\cite{perth97}), U.C. Berkeley (\cite{KAIT}), and by R. Evans
(\cite{evansref}).  These searches fill in the distant and nearby ends
of the Hubble diagram, respectively. Though many SNe have been found
at $0.002 < z < 0.2$, selection effects plague most samples, and are
nearly impossible to quantify. A well-understood, complete sample of
SNe in this intermediate distance range is needed so that we can
understand the uncertainties and biases which arise from calibrating
the distant samples using the nearby objects. The search described in
this paper is optimized for finding SNe over the range of redshifts
$0.02\leq z\leq0.08$ with a completeness limit fainter than $V>20$.
Complimentary searches scanning lower-redshift Abell clusters
(\cite{maza97}) and Northern hemisphere Abell clusters
(\cite{utexas97}) have also been initiated.  The SNe produced by these
searches, when combined with objects from the nearby and distant
searches, will comprise a sample of SNe encompassing a broad range of
distances with which we can accurately map out the expansion of the
Universe from $0.002\leq z\leq 1.0$.

To understand the systematics involved in using SNe as distance
indicators, we need to understand the SNe, their evolution, and their
progenitors better. Supernova rates, for example, place important
constraints on models of progenitor evolution and the physical
processes involved in the explosions, in addition to star formation
and chemical enrichment. However, they are still subject to dispute,
particularly due to interpretations of selection effects and control
times and the large statistical uncertainties of small samples (see
\cite{cappellaroetal94}, \cite{tammannrates94}). Our sample will be
unique in that it is volume-limited to more than a magnitude below the
SN~Ia average brightness for our most distant clusters, and its
selection criteria are well understood. These properties will allow us
to estimate and compare rates of SNe of different types, both in the
clusters, and in the field surrounding them.

Knowledge of the luminosity distribution of SNe~Ia would enable us to
place constraints on the systematics (such as Malmquist bias)
introduced by using SNe~Ia as distance indicators. However, the SN~Ia
luminosity function is largely unknown due to poor understanding of
the selection effects of past searches. An important aspect of this is
the evolution of the luminosities of SNe~Ia.  \cite{hetal96a} have
shown that there appears to be an intrinsic difference in the peak
luminosity--decline-rate relationship between the nearby \cite{phil93}
sample and their more distant sample of SNe~Ia. In addition, there
appears to be a strong correlation between the decline rate of SNe~Ia
and the host galaxy type; SNe~Ia that occur in early-type galaxies
have preferentially narrower light curves than those in their spiral
counterparts.  Fortunately, after correcting for light curve shape,
there is no detectable differences between distances measured to
early- and late-type galaxies (\cite{schmidt97}). Our SN sample should
allow us to construct a SN~Ia luminosity function and to explore the
effects of the host galaxy type, progenitor stellar population ages,
and environment (cluster vs. field) on SN luminosities. This will
enable us to place limits on these evolutionary effects and to
estimate any biases which they might introduce into estimates of
$q_{0}$ via SNe Ia.

The intermediate-distance SNe are also ideally suited for measuring
the reflex motion of the Local Group (LG), unhindered by random
peculiar velocities of nearby galaxies or large uncertainties in
measuring distances of very distant objects. Lauer and Postman (1994)
[LP94] \nocite{lp94} determined distances to Abell clusters within
$z<0.05$ using brightest cluster galaxies (BCG) as distance indicators
with an accuracy of 16\%. Their controversial result suggested that
these clusters participate in a large bulk motion (560~km~sec$^{-1}$)
pointing nearly $70^{\circ}$ away from the direction of the dipole in
the microwave background (CMB) as measured by COBE (\cite{CMBref}).
The amplitude of their measurement within such a large volume
contradicts many currently popular models (\cite{feldman+watkins94},
\cite{strauss+etal95}). While other studies of Abell clusters
(\cite{hudson+ebeling97}, \cite{branchini+etal96}) have questioned the
significance of the LP result, other reanalyses (\cite{colless95},
\cite{graham96}) have supported their claims, and simulations have
confirmed the notion that dense clusters can readily be used to trace
the large-scale structure of the universe (\cite{gramann+etal95}).

Studies of large-scale motions using other distance estimators have
since been attempted to test the validity of the LP result.  In
particular, \cite{rpk95} [RPK95] used a sample of 13 SNe~Ia to measure
peculiar velocities to galaxies in the field. Using the uncertainties
in the SN~Ia distance measurements which are computed directly from
their multi-color light-curve shapes technique, they found that they
can rule out the LP measurement at a confidence level better than
99.3\%.  Moreover, their velocity estimate is consistent with a bulk
motion of similar amplitude and direction to the velocity dipole in
the CMB.  Still other peculiar velocity samples constructed using
different distance indicators, such as Tully-Fisher measurements
(\cite{giovetal96}), also seem to disagree with LP94. However,
\cite{watkins+feldman95} argued that in contrast to the BCG sample of
LP94, these more recent measurements are sensitive to small-scale
motions, to different degrees (since motions of rich galaxy clusters
are less sensitive to small-scale flows) and this significant source
of noise reduces their overall sensitivity to the large-scale motions.
Therefore, because these samples probe different locations in space
where the small-scale flows differ, this results in decreasing, but
not eliminating, the significance of the disagreement between the bulk
motion measurements of LP94 and RPK95 (and by extension the
Tully-Fisher measurements of \cite{giovetal96}).

The Abell cluster targets in this search are a subset of the clusters
currently being observed as part of a larger BCG survey by Postman,
Lauer and Strauss (PLS\nocite{pls}). The resulting sample of cluster
peculiar velocities from SNe~Ia will be directly comparable to the BCG
measurements which will emerge from the LP94 + PLS surveys, allowing a
direct comparison of SN~Ia distances and BCG distances to many
clusters. This will enable us to measure the reflex motion of the LG
with respect to a subset of the LP94 + PLS clusters using an
independent distance indicator which offers significantly increased
precision, and which can be directly compared to that of LP94 and/or
PLS, in contrast to the other surveys mentioned above.  The
contribution that this SN search can make to the bulk motion question
is examined in detail in Section~\ref{sec:dipolesims}.  Finally, the
large-scale velocity field of Abell clusters derived from our sample
of SNe can be compared to those predicted by the gravity fields of
all-sky galaxy catalogs and simulations, to test models of structure
formation and constrain $\Omega_{0}^{0.6}/b$, where $b$ is the biasing
parameter, similar to the analysis performed on a sample of 25 SNe~Ia
in the field by \cite{riess+davis97}.

Additional projects of interest which may emerge from this project
include: (1) placing limits of low-mass intracluster MaCHOs ($\sim
10^{-4}M_{\sun}$) by trying to detect microlensing light curves on top
of the light curves of SNe which are found on the far side of their
cluster fields (\cite{kolatt+bartelmann97}), (2) detecting the tidal
disruption and accretion of stars into the central black holes of AGN
through flares in the AGN luminosity (\cite{agn_flare89}), and (3)
detecting RR-Lyrae in the halo of our galaxy, useful for tracing its
kinematics.

\section{Search Strategy}
\label{sec:strategy}
The difficulty in searching for intermediate-distance supernovae is
that one cannot practically target individual galaxies with short
exposure times as in nearby searches, while the volumes (and galaxy
numbers) being sampled are much smaller than those in the
high-redshift searches.  One must survey a large area of sky at
locations where the galaxy number density is the largest at the
desired distance, using an intermediate-sized telescope. The Mount
Stromlo Abell Cluster Supernova Search employs the MaCHO wide-field
($45^{\prime}\times 45^{\prime}$) dual-color mosaic imager
(\cite{cameraref}) on the MSSSO 1.3m telescope (\cite{machoref}). The
camera's passbands are not standard and we refer to them as $V_{M}$
for the `blue' side and $R_{M}$ for the red side. We specifically
target all high-density ($N_{gal} \equiv$ number of galaxies within
one Abell radius $\geq 65$) clusters from the Abell, Colowin and Owen
(ACO, 1989) \nocite{aco89} survey of nearby, rich galaxy clusters.
The target clusters are further constrained to lie within $z\leq 0.08$
as measured in the BCG survey of PLS, \nocite{pls} and to be
accessible at reasonable altitude from Stromlo ($\delta (J2000)\leq
+15$). The complete sample of 74 clusters is listed in
Table~\ref{tab:targets}; their distribution in the sky is displayed in
Figure~\ref{fig:skydist}. In all there are 20 clusters in our sample
whose BCG distances were measured by LP94. Note that the sample's
large-scale distribution on the sky is not uniform; in particular
nearly 1/6 of the target clusters are located in the direction of the
Shapely-Ames Supercluster near $(l,b)\sim (315,+30)$.

\begin{table}
  \dummytable\label{tab:targets}
\end{table}

Our project has been allocated 5\% of the telescope time on the 1.3m,
the remaining 95\% belonging to the MaCHO project, with all of our
images being observed by the MaCHO staff observer. A typical SN~Ia at
$z=0.08$ has $M_{B}\sim 18.5$ near maximum light. Type~II and Ib/c SNe
are 0 to 5 magnitudes fainter. In order to be complete in detecting
point sources at 20th magnitude, the faintest objects which we can
readily follow up with photometry and spectroscopy, with
$2.5^{\prime\prime}$ FWHM seeing that is typical at MSO
(Figure~\ref{fig:seeinghist} shows the distribution of observed
FWHM over the past year), we take 240-second exposures of each field.
Therefore we have enough time to average $\sim 4$ to 5 observations
night$^{-1}$. The telescope time is most efficiently used (with less
overhead) by observing $\sim 14$ fields every third night.  Allowing
for weather ($\sim 50\%$), we average $\sim 14$ fields per week,
enough to observe $\sim 1/3$ of the 74 fields in the two-week period
that it takes for a SN~Ia to reach peak brightness.

The Macho camera's pixel scale of $0.628^{\prime\prime}/$pixel,
combined with the $\sim 2.5^{\prime\prime}$ typical FWHM, allows for
good sampling of the PSF, which is important in the image subtraction
process used to detect SNe, as described in \S\ref{sec:searching}. The
dual mosaic focal plane is read out at 16 separate amplifiers,
resulting in 15 separate $1024\times 2048$-pixel images (8 in blue and
7 in red because of one dead amplifier) which sample the entire
$45^{\prime}$ field. This also is advantageous for the subtraction
process, as will be explained below.

High-quality and frequent followup photometric and spectroscopic
observations are extremely important for the success of this project.
In addition to the images from the 1.3m, \footnote{The non-standard
  V$_{M}$ and R$_{M}$ passbands can be accurately transformed to
  rest-frame B and R$_{C}$ (0.01~mag r.m.s.)  as described by
  \cite{kimetal96}, because they are well matched to these filters
  redshifted to $z\sim 0.06$ (\cite{paperIII}).} we obtain regularly
scheduled observations on the MSSSO 2.3m (spectra and imaging), and
the MSSSO 1.9m. The ARC 3.5m is also employed for spectroscopic
followup on SNe detected at $\delta\simgt -30$. In addition, we use a
semi-dedicated (shared with the MaCHO group) $30^{\prime\prime}$ at
MSSSO as our main source of photometric coverage. The nightly
observations are conducted by the RAPT group of amateur astronomers
and observers, who are an invaluable asset to this program.
\footnote{For more information on the RAPT group, please see
  http://www.tip.net.au/\~{ }bnc/.}

\section{Scheduling and Data Collection}
\label{sec:scheduling}
An automated scheduling routine is used to determine which fields (of
those that satisfy the criteria described in \S\ref{sec:strategy}) are
best for observing on a given night, and at what time. The algorithm
is summarized below:

\begin{enumerate}
\item A field is ineligible for observing on any given night if:
  \begin {enumerate}
  \item its hour angle at evening twilight is HA$_{eve}>2$ hours, or
  \item its hour angle at morning twilight is HA$_{morn}<-4$ hours.
    This ensures that the clusters will be observable for at least 6
    weeks (allowing sufficient followup).
  \end{enumerate}
\item Assign a weight $w_{i}$ (which is a function of time, $t$) to
  each of the remaining fields, $i$:
  \begin{equation}
    w_{i}(t) = \left(\frac{N_{days,i}}{21}\right)^2 \times \frac{N_{gal,i}}{100} 
    \times \frac{X_{i}(t)}{X_{min,i}}, \label{eq:weight}
  \end{equation}
  where $N_{days,i}$ is the number of days since the $i^{th}$ field's last observation, 
  $N_{gal,i}$ is its ACO galaxy count, $X_{i}(t)$ is its current airmass, and $X_{min,i}$ 
  is the minimum airmass which it reaches at Mount Stromlo. 
  \begin{enumerate}
  \item Special consideration is taken for fields which are being
    observed for the first time this season so that they don't
    dominate the schedule.
  \item If the night is dark and the seeing good (better than
    $2^{\prime\prime}$), weights for fields which have not yet been
    observed are artificially inflated to allow more template
    observations on good nights (see \S\ref{subsec:template}).
  \item Weights for fields which have been observed within the past
    week are artificially reduced so they cannot be observed at the
    expense of other fields.
  \item Individual clusters can be given higher priority, as desired,
    for example, to increase the frequency of observations made of
    clusters with current SNe, or those with possible SNe which need
    confirmation.
  \end{enumerate}
\item \label{item:findtc} The time, $t_{c}$, which maximizes the sum
  of weights of all observable fields, \(\sum_{i} w_{i}(t)\), is
  chosen as the best time to observe.
\item Queue the fields which are observable at time $t_{c}$ by their
  weight~$w_{i}(t_{c})$, filling up the allocated 15\% of the night
  (including time for CCD readout and telescope slewing).
  \footnote{Though steps \protect\ref{item:skipnearmoon} and
    \protect\ref{item:skiptoolow} logically should come prior to step
    \protect\ref{item:findtc}, our scheduling follows this sequence so
    that we do not use more than our 5\% share of dark time on the
    1.3m. There are usually sufficient numbers of clusters needing
    observations that this procedure does not result in any wasted
    observing time.}
  \begin{enumerate}
  \item \label{item:skipnearmoon} Skip those clusters which lie less
    than $20^{\circ}$ from the moon.
  \item \label{item:skiptoolow} Also skip those clusters which lie at
    $X_{i}(t_{c})>2.0$.
  \end{enumerate}
\item Sort the scheduled fields by their azimuthal angle to provide
  the telescope with the shortest slewing path between observations.
\end{enumerate}

The routine creates a `scheduler file' which lists the coordinates of
the scheduled fields; this is fed directly into the MaCHO
camera/telescope controller system at the requested time, $t_{c}$.
The automated system takes over at that point until all exposures are
finished. The resulting bias-subtracted, flat-fielded images are then
copied to local disk for analysis.

The amount by which each factor in Eq.~\ref{eq:weight} contributes to
the cluster weights has been tuned so that on average, each cluster is
revisited every fortnight -- even those with the lowest galaxy counts
in our sample. This system has worked extremely well; to date $\sim
4.7\%$ of the usable telescope time has been used to observe Abell
clusters. As shown in Figure~\ref{fig:schedhist} (top), for the
observations of the past year, the median time-between-observation
($\Delta t$) of each field is 12 to 15 days, with a strongly peaked
distribution (this histogram also includes the observations taken
during the first few months of the search, in which the scheduler was
being `tuned'). A very similar distribution is obtained from a
simulation of 3 years' observations in which a random 50\% of the
nights are skipped (Fig.~\ref{fig:schedhist}, bottom).  It is evident
from Fig.~\ref{fig:schedhist} that a number of clusters do not get
observed for 3 weeks or more, but this is to be expected when random
weather patterns, the moon, scheduled followup, and candidate object
re-observations, etc. are considered. We have recently introduced
other measures which we expect will decrease the large number (though
still $<10\%$) of observations for which $\Delta t = 3$ days. Still,
80\% of the observations were made with $\Delta t \leq 21$ days, and
although sparser clusters do slip through the cracks more easily, as
Fig.~\ref{fig:schedhist} demonstrates, the amount of favor given to
larger clusters is small.  The overall results are more than adequate,
and we believe the scheduling could not be substantially improved by a
human counterpart.

\section{Searching for Supernovae}
\label{sec:searching}
The software which searches the images for supernovae consists of a
series of IRAF tasks and programs written in C, linked together and
automated via a set of scripts written in PERL, with final cuts on
potential discoveries being made interactively. Briefly, it involves
subtracting an earlier observation of the field (a {\it template})
from the observation so that all objects which have increased in flux
appear as new point-sources. No attempt is made to combine the 8
amplifier images for each color into one $4096\times 4096$ image prior
to searching; instead each amplifier is processed individually. This
has several advantages and one disadvantage, which will be pointed out
below.  Figure~\ref{fig:searching} illustrates the method, which can
be summarized by the following expression:
\begin{equation}
{\cal S} = b \times \left( a + {\cal O}\ast {\cal K} \right) - {\cal T},
\end{equation}
where the subtracted image, ${\cal S}$, is scanned for new objects.
Here, ${\cal T}$ is the template of the field, ${\cal O}$ is the
observation (registered to the template), ${\cal K}$ is a convolution
kernel required to match the PSF of ${\cal O}$ with that of ${\cal
  T}$, and $a$ and $b$ are an additive constant and linear scale
factor, respectively, which match the background and flux of objects
in ${\cal O}$ to those in ${\cal T}$. The following sections describe
the algorithm in detail.

\subsection{Constructing the Template}
\label{subsec:template}
Before any image subtraction can be done, a template of the field must
be constructed.  This involves (1) preprocessing -- masking of
saturated stars and bright pixels and removing any linear gradients in
sky brightness (all observations are pre-processed identically); (2)
detecting bright stars for image alignment (we use DoPhot,
\cite{dophotref}, which also provides an analytic model of the stellar
PSF); (3) identifying isolated high S/N stars of differing brightness
for use in PSF and flux matching, and (4) laying down a grid of
(negative) false stars of known flux for quick assessment of searching
depth after the template is subtracted from the observation. The
template is made using an observation obtained preferably during a
dark, transparent night with good seeing. If a later observation
proves to be better than the template, the template is replaced after
that observation has been searched.  This is a simple procedure since
the process is completely automated.  A section of an example template
is shown in Figure~\ref{fig:searching}a.

\subsection{Image Registration}
\label{subsec:registration}
The most important aspect of the image subtraction process is
accurately aligning the observation images to the template images.
The registration is accomplished using a triangle-matching algorithm
(see \cite{triangles}) to determine the coordinate transform between
bright stars found using DoPhot in the two images. In our case since
the template and observations are always taken on the same
telescope/camera system, we can constrain the rotation and scale in
the transform, which allows a more robust convergence with fewer
failures. The transform is then entered into the IRAF GEOTRAN routine
which performs the flux-conserving linear geometric transform of the
observation image, using a linear interpolation between pixels for
subsampling. Since the PSF is well sampled, resampling errors are
small. The registration typically requires a shift of $<50$ pixels and
is accurate to $\sim 0.3$ pixels across an entire individual amplifier
image. This process is aided by not registering the entire
$45^{\prime}$ field at once; the effects of variations in pixel scale
across the field are reduced and higher-order transformations rendered
unnecessary. However, it does result in a loss of $\sim 50$ pixels of
searchable area near the edges of each amplifier image after the
registration. In Figure~\ref{fig:searching}b we present a portion of an
observation before (left) and after (right) it is registered with the
template of Figure~\ref{fig:searching}a.

\subsection{PSF Matching}
\label{subsec:psfmatching}
If the observation and template were both photometric and their PSFs
were identical, a simple subtraction would now be suitable.
Unfortunately, this is almost never the case. Instead, we must match
the images' PSFs and intensities.  For the sake of simplicity, we will
assume that we are convolving the template, since the template is
usually the better image, and the better of the two images is the one
which must be degraded to match the poorer one. The convolution kernel
which can be used to convolve the template so that its PSF matches
that of the registered observation is computed using Andrew Phillips'
implementation of the algorithm of \cite{ciardullo90} (see
\cite{alinear95}) in IRAF (called QPSF, in the ALINEAR package).
Briefly, the method computes the kernel in Fourier space, where a
convolution is a simple multiplication. Since the PSF is very nearly
Gaussian, its Fourier Transform (FT), and thus the convolution kernel,
will also be very nearly Gaussian. However the high-frequency
components of the FT become dominated by noise in the wings of the PSF
where the signal becomes weakest. Still, one can model the high S/N
components of the FT with an elliptical Gaussian, and the wings of
such a model can be used in place of the noise-dominated components.
Thus a single, bright, isolated star can be used to compute the
required convolution kernel. In the case where the two images are
similar (FWHMs differ by $<0.3$ pixels), the convolution kernel may be
very poorly resolved, or even undefined (as might be the case if the
PSFs are elongated). One can overcome this problem by degrading the
slightly poorer image using a Gaussian convolution kernel before
performing the PSF matching. The Fourier method is applied to each of
the individual amplifier images, sufficiently reducing the effects of
any small-scale PSF variation across the focal plane.  The template of
Figure~\ref{fig:searching}a is shown in Figure~\ref{fig:searching}d
after it has been convolved to match the registered image of
Figure~\ref{fig:searching}c.

\subsection{Photometric Scaling and Subtraction}
\label{subsec:scaling}
For the photometric scaling, as in the PSF matching, all that is
required is a high S/N, isolated star or galaxy. The IRAF ALINEAR
task, ITRAN (\cite{alinear95}), performs a linear least-$\chi^{2}$ fit
to the difference, in ADU, between the pixels which surround the
chosen bright object in a subsection of the two images (registered
observation and convolved template). The resulting slope (measuring
the ratio in flux between the two images) and offset (measuring the
difference in sky brightness) can then be applied to the registered
observation. The slope, offset, and $\chi^{2}$ give us further
information on the quality of the observation and subsequent
subtraction. For example, the star chosen for PSF-matching might be
saturated in the observation, say, if the seeing was extremely good,
and the $\chi^{2}$ will be extremely high. In this case we can go back
and match the PSFs and fluxes using a slightly fainter star and check
the results. Again it is worth noting that having the focal plane
divided into eight smaller images makes the scaling more accurate as
any large-scale nonlinear gradients in sky brightness or scattered
light (such as due to the moon) or even variation in transparency
across the $45^{\prime}$ field (due to thin clouds) are reduced.
It makes even more sense when one considers that the individual images
come from different CCDs and thus have different gains, bias levels,
and color terms.

Finally, the intensity-transformed, convolved template now reflects
the same atmospheric and photometric conditions as the registered
observation, and a simple subtraction of the first from the second can
be performed. The subtraction of the convolved template image
(Figure~\ref{fig:searching}d), after being photometrically scaled,
from the registered observation (Figure~\ref{fig:searching}c) is
presented in Figure~\ref{fig:searching}e. Here, a supernova is clearly
visible where in the original observation
(Figure~\ref{fig:searching}b), it was hidden in the background light
of its host galaxy. Five of the (now positive) sensitivity stars are
also visible in the subtracted image. The corresponding magnitude
limit for this image is $R_{M}\sim 20.5$ (typical); the supernova is
$\sim 0.9$ magnitudes brighter. Note that the noise in the subtracted
image is now a function of Poisson noise (from the shifted
observation) plus correlated Poisson noise (from the convolved
template) plus residual noise left over from the subtraction.

\subsection{Object Detection}
\label{subsec:detecting}
Once the subtraction is complete, a new supernova will appear as a
point source in the subtracted image, with a FWHM which is the same
size as other stars in the original observation. To find these, we use
a point-source detection algorithm, written in C, which samples the
subtracted image at many locations over the scale of the PSF to
estimate the total noise ($\sigma$) within one correlation length. It
then detects objects with a total flux that is 3 $\sigma$ above the
background. If a detection lies at the same location as a bright star
on the template, it is assumed to be a poor subtraction, and is
eliminated from the list.  (Poor subtractions near bright stars are
common because the large flux gradients magnify the effects of even a
small registration or PSF-matching error; such a residual can be seen
above and to the left of the supernova in Fig.~\ref{fig:searching}e.)
In addition, our routine tests if each detected object is consistent
with a point source of given peak intensity and known FWHM. This helps
eliminate cosmic rays, which resemble stars (since they have been
convolved) but are slightly narrower than true stars in the image.
Often a poor subtraction results in an area of negative flux next to a
detected object. Our algorithm tests for these as well and eliminates
any such suspicious objects. The routine then eliminates groups of
objects which fall in straight lines (satellite trails or bad
columns). As a check on the depth and robustness of the search, the
false sensitivity stars which have been detected are identified, and
the corresponding limiting magnitude is computed from the known flux
of the faintest star detected.

Though the object-detection algorithm has been custom designed to
reduce the number of false detections, several hundred usually pass
the cuts for each observation. Most of these offenders are cosmic
rays; thus, our ultimate defense is that we have two observations made
simultaneously, one in $V_{M}$ and one in $R_{M}$. As a final check,
the locations and fluxes of the objects detected in the $V_{M}$ images
are compared to those in the $R_{M}$ images. This serves to cull
nearly all cosmic ray hits. In the end, our detections are $\sim$4.5
$\sigma$ above the correlated noise (recall this is {\it not} the
Poisson noise of the original image, but can be similar to it in a
good subtraction, or significantly greater than it in a poor
subtraction). On average 5--15 objects per observation remain, and
these comprise poor subtractions near bright galaxies and stars
(somewhat common, particularly in poor seeing); cosmic rays which
coincidentally lie at the same locations in both color images (rare);
and true astronomical objects ($>95\%$ of which are asteroids). For
each of these objects, a subraster image is created, and final cuts
are made by eye.

\subsection{Subrasters and Object Classification}
\label{subsec:subrasters}
The layout of a subraster image produced by the search software is
shown in Figure~\ref{fig:subraster}, in this case, containing the SN
seen in Figure~\ref{fig:searching}e. The subrasters allow for quick
examination, by including the candidate object in both colors ($V_{M}$
on the left, $R_{M}$ on the right), in the template (bottom),
observation (top) and subtraction (center). They are archived on disk
for quick retrieval and can be accessed at any time by running a
script which displays the subrasters, and allows interesting objects
to be scrutinized in detail using IMEXAM in a local XIMTOOL window.
Alternatively, they can be viewed over the World Wide Web on an
interactive web page which displays and then IMEXAMs them in a local
XIMTOOL window, and allows object classifications to be recorded in a
database for comparison to future observations (see below). Both
methods also display the object's coordinates and approximate
magnitudes in both colors. In addition, information on the quality of
the observation, such as the scale and offset used in flux-matching,
the FWHM, and the number of sensitivity stars detected, are all
reported to the classifier.  All of these data are used to quickly
classify objects among poor subtractions and cosmic rays, asteroids,
variable stars which have increased in brightness, and supernova
candidates. We find that the human eye is extremely efficient at
classifying the objects. A great deal of experience is gained by
examining subrasters day after day, and a certain amount of intuition
certainly plays a role in object classification; however certain
concrete rules also apply.

SNe, variable stars, and asteroids each have characteristics which aid
in their identification, and the color information which we have is
extremely helpful. Young supernovae are bluish objects ($V_{M}-R_{M}<
0.2$), and are usually associated with a galaxy.  If an object lies
directly on the nucleus of the galaxy, it is treated with suspicion
and is often a poor subtraction or AGN; however we have detected
several SNe very near the centers of their host galaxies, SN 1997bz
being one example.  Asteroids, on the other hand, are usually redder
than the color of the Sun ($V_{M}-R_{M} > 0.3$), are usually not
associated with a galaxy, and upon close inspection often exhibit
slightly elongated PSFs (particularly the bright ones).  Unfortunately
there is a large dispersion ($\sim 0.5$ magnitudes) in the rough
uncalibrated $V_{M}-R_{M}$ colors computed by the software, thus color
discrimination provides an aid but not an end in object
classification. Variable stars are straightforward to identify as they
are present as a star on the template image, as well as in the
observation.

The classification-by-eye scheme is certainly not fail-proof (it is
possible that we do misidentify some SNe as asteroids). We overcome
this weakness by keeping records of the positions of all objects which
we identify. If an object is identified as an asteroid, we can compare
its location to those of other objects identified in a later
observation of the same field. If it is present at the same location
in both observations, it is likely to be a SN rather than an asteroid.
On the other hand, any field with a probable supernova as identified
by eye is re-observed as soon as possible to see if the object has
moved or disappeared; if it has, it is most likely an asteroid. We
perform such followup on all objects located near galaxies, regardless
of color, and on blue objects not associated with galaxies.  In the
end, this conservative approach eliminates all possibility of
misidentifying an asteroid as a supernova.  Our probability of
misidentifying a supernova as an asteroid is not zero, but if we
re-observe the field within the next few weeks, we are able to correct
such misidentifications. However we will still have a diminished
sensitivity to SNe with undetected hosts.

\section{Results}
\label{sec:results}
Once a promising supernova candidate is found, and has also been
confirmed with a second observation, it is announced in an IAU
Circular.  A finder chart with positions accurate to better than
$0.5^{\prime\prime}$ is immediately created using an automated
routine, and all available information is placed on our web page.
Information on the 19 supernovae discovered by this search to date are
listed in Table~\ref{tab:discoveries} (we came on line in early June,
1996).  Although spectroscopic followup was sparse at the beginning
due to weather and technical difficulties, we anticipate being able to
get a redshift and spectral type for nearly all future supernovae. The
rate of discovery is about 1.5 month$^{-1}$. Of the 15 SNe for which
redshifts have been measured, 8 appear to be associated with their
host Abell cluster, while six lie behind their target clusters, and
one is in a foreground galaxy. Six of the nine SNe which have been
spectrally classified are SNe~Ia. Many of the other SNe have occurred
in elliptical galaxies and can be identified as type~Ia on that basis
alone. We can also, with reasonable reliability, classify SNe using
their multicolor light curves. In total, we find that 12 objects, of
the 17 which have spectra and/or reduced light curves so far, are SNe
$\sim$Ia, and 7 of these are associated with their host Abell
clusters.  Two classified SNe are type~II and are associated with
their host clusters, while one other, classified as type~II, is not.
The two remaining objects are probably not type~Ia, and/or they are
not associated with their host clusters. Light curves, classification,
and analysis will be presented in an upcoming paper (\cite{paperIII}).

We keep records of the limiting magnitudes resulting from each
observation. The distribution for observations over the past year are
shown in Figure~\ref{fig:mlimhist}. The median limiting search
magnitudes are $\sim 20.4$ in $R_{M}$ and $\sim 20.5$ in $V_{M}$,
though we expect these limits to get fainter as we accumulate better
templates. The large dispersion can be attributed to variation in sky
conditions. A more complete discussion of our magnitude limits and
their effects on the completeness of our search, as well as
implications for the SN rate, will be presented in a paper currently
in preparation (\cite{paperII}).

\begin{table}
  \dummytable\label{tab:discoveries}
\end{table}

\section{Determining the Motion of the Local Group}
\label{sec:dipolesims}
In this section we address how well we expect to be able to determine
the motion of the Local Group (LG) relative to the Abell Clusters in
our sample using SNe~Ia. To do this, we have run a series of Monte
Carlo simulations in which 10,000 fake samples have been created. The
samples have varying sizes, geometries (in the sky), and bulk motion
vectors. For the following discussion, all velocities mentioned are
relative to the LG frame.  In each case, the `fake' supernovae have
been given typical peculiar velocities of 400 km~sec$^{-1}$
(\cite{marzke95}) on top of the flow velocity, distance uncertainties
of 8\%, and redshift uncertainties of 0.001 (in this redshift range,
the overall errors are dominated by the distance uncertainties). The
peculiar velocities which we are to use for our measurement are likely
to be smaller ($\sim 250$ km~sec$^{-1}$) because we can use the
average peculiar velocities of several galaxies in the host clusters
for many of our SNe; however, note that we only include uncorrelated
motions, and neglect small-scale correlated motions described in
\cite{watkins+feldman95}, which Abell clusters are less sensitive to
than galaxies in the field. Still, the uncertainties are dominated by
the distance uncertainties (a distance uncertainty of 8\% at z=0.08 is
a $\sim 1900$ km~sec$^{-1}$ velocity uncertainty), so these
simulations do not present unrealistic errors (this was verified by
running simulations using different uncertainties and peculiar
velocities).  The redshift distribution of the SNe in the simulations
mimics that of our cluster sample. Once the simulated SN sample is
created, the best-fit values for the Local Group velocity vector,
$\vec{v}_{LG}$, and the Hubble Constant, $H_{0}$, for this sample are
determined simultaneously by minimizing the statistic
\begin{equation}
\chi^{2} = \sum_{i} \frac{\left( c z_{i} - H_{0} d_{i} + \vec{v}_{LG}
    \cdot \hat{r}_{i} \right)^{2}}{\sigma_{i}^{2}} ,
\end{equation}
where $d_{i}$ is the distance of the $i$th supernova, $z_{i}$ is its
redshift, $\hat{r}_{i}$ is the unit vector pointing in its direction
in the sky, and $\sigma_{i}^{2}$ is the quadrature sum of all errors
mentioned above.

Of primary interest is estimating the accuracy with which our sample
can be used to measure the reflex motion of the LG.  To answer this,
we compute the 1-$\sigma$ dispersions in the 10,000 simulated dipole
measurements for a sample geometry which matches that of our cluster
targets. These are presented in Table~\ref{tab:simresults} for the 3
velocity components, for a 20-, 40-, and 60-SN sample, using the
parameters described above.  The dispersions are identical for all
sample geometries which we explored, and are independent of the
direction of the input velocity vector.  By blindly comparing these
uncertainties to those quoted by LP in their 1994 measurement,
($\pm$250, $\pm$273, $\pm$198) km sec$^{-1}$, we find that applying
$\sim$35 SNe~Ia distances collected in our sample to the problem
results in a comparable measurement to that made using the 124 BCG
distances of the LP sample. This is not surprising noting that SNe~Ia
are $\sim 2\times$ more accurate as distance indicators.  For example,
were we to make a velocity measurement, using the SNe~Ia in our
sample, that is consistent with the LP result, it would be significant
at 2.3 $\sigma$, 3.7 $\sigma$, and 4.9 $\sigma$ levels for 20, 40 and
60 SNe~Ia, respectively, compared to the COBE measurement.  On the
other hand, a measurement which coincides with the COBE dipole would
rule out the LP result at the 1.9 $\sigma$, 2.6 $\sigma$, and 2.9
$\sigma$ levels (using the uncertainties of LP94).  Note, however,
that the simulated distance uncertainties of 8\% are slightly greater
than that estimated by \cite{rpk96} for their well-sampled SNe~Ia
light curves; this along with the high velocity dispersions used in
the simulations, as pointed out above, implies that the results
described here are probably slightly conservative. A comparison of the
uncertainties listed in Table~\ref{tab:simresults} for 20 SNe with
those quoted by RPK95 (for 13 SNe) of ($\pm$370, $\pm$510, $\pm$220)
confirms this notion.

In addition to estimating the expected uncertainties, the simulations
also allow us to confirm that there will be a small geometric bias in
any dipole measurement which is made using our sample, due to its
non-uniform geometry (see Fig.~\ref{fig:skydist}). By comparing the
fits resulting from the simulated samples to their input dipole
velocity vectors, we find that there will likely be a $\sim \pm 5$ km
sec$^{-1}$ offset in each vector component from the true value. The
value will of course depend upon the actual geometry of the sample and
the measured bulk motion, and can be easily computed and corrected,
but it is still reassuring to find that it will be small compared to
the expected uncertainties. We will investigate this, and other
possible sources of systematic error, in a subsequent paper on the
distance measurements from our first year's set of SNe.

\begin{table}
  \dummytable\label{tab:simresults}
\end{table}

\acknowledgements{ We would like to thank the observatory director at
  MSSSO for his selection of this project for director's discretionary
  time. We are also grateful to M.~Postman, T.~Lauer, and M.~Strauss
  for donating their compilation of cluster data for our target
  selection, and to A.~Riess for his insightful comments on the
  manuscript. This research is supported by NSF grant AST~9617036 and
  by grants from the Seaver Institute and the Packard Foundation to
  the University of Washington. }

\newpage


\begin{thebibliography}{}

\bibitem[Abell, Corwin, \& Olowin 1989]{aco89}
Abell, G.~O., Corwin, H.~G., J., \& Olowin, R.~P. 1989, { Astrophys. J. Supp.
  Series}, {\rm 70}, 1.

\bibitem[Adams \etal  1997]{utexas97}
Adams, M.~T., Howell, D.~A., Ward, M.~H., Wheeler, J.~C., \& Wren, W. 1997,
  {\it IAU Circ \#6674}.

\bibitem[{Branchini}, {Plionis}, \& {Sciama} 1996]{branchini+etal96}
{Branchini}, E., {Plionis}, M., \& {Sciama}, D.~W. 1996, { Astrophys. J.}, {\rm
  461}, L17.

\bibitem[Cappellaro \etal  1993]{cappellaroetal94}
Cappellaro, E., Turatto, M., Benetti, S., Tsvetkov, D.~Y., Bartunov, O.~S., \&
  Makarova, I.~N. 1993, { Astr. Astrophys.}, {\rm 268}, 472.

\bibitem[Ciardullo, Tamblyn, \& Phillips 1990]{ciardullo90}
Ciardullo, R., Tamblyn, P., \& Phillips, A.~C. 1990, { Publ. Astr. Soc.
  Pacific}, {\rm 1113}, 102.

\bibitem[{Colless} 1995]{colless95}
{Colless}, M. 1995, { Astron. J.}, {\rm 109}, 1937.

\bibitem[d.~Li \etal  1997]{baoref}
d.~Li, W., l.~Qiu, Y., y.~Qiao, Q., Zhang, Y., Zhou, W., \& y.~Hu, J. 1997,
  {\it IAU Circ \#6661}.

\bibitem[{Evans} \& {Kochanek} 1989]{agn_flare89}
{Evans}, C.~R. \& {Kochanek}, C.~S. 1989, { Astrophys. J.}, {\rm 346}, L13.

\bibitem[Evans 1997]{evansref}
Evans, R. 1997, {\it IAU Circ \#6613}.

\bibitem[Feldman \& Watkins 1994]{feldman+watkins94}
Feldman, H.~A. \& Watkins, R. 1994, { Astrophys. J.}, {\rm 430}, L17.

\bibitem[Fixsen \etal  1994]{CMBref}
Fixsen, D.~J. \etal  1994, { Astrophys. J.}, {\rm 420}, 445.

\bibitem[Germany \etal  1997]{paperIII}
Germany, L., Reiss, D., Schmidt, B.~P., \& Stubbs, C.~W. 1997, {\it In
  Preparation}.

\bibitem[Giovanelli \etal  1996]{giovetal96}
Giovanelli, R., Haynes, M.~P., Wegner, G., Da~Costa, L.~N., Freudling, W., \&
  Salzer, J.~J. 1996, { Astrophys. J.}, {\rm 464}, L99.

\bibitem[{Graham} 1996]{graham96}
{Graham}, A.~W. 1996, { Astrophys. J.}, {\rm 459}, 27.

\bibitem[{Gramann} \etal  1995]{gramann+etal95}
{Gramann}, M., {Bahcall}, N.~A., {Cen}, R., \& {Gott}, J.~R. 1995, { Astrophys.
  J.}, {\rm 441}, 449.

\bibitem[Groth 1986]{triangles}
Groth, E.~J. 1986, { Astron. J.}, {\rm 91}, 1244.

\bibitem[Hamuy \etal  1995]{hetal95}
Hamuy, M. \etal  1995, { Astron. J.}, {\rm 109}, 1.

\bibitem[Hamuy \etal  1996a]{hetal96a}
Hamuy, M. \etal  1996a, { Astron. J.}, {\rm 112}, 2391.

\bibitem[Hamuy \etal  1996b]{hetal96b}
Hamuy, M. \etal  1996b, { Astron. J.}, {\rm 112}, 2408.

\bibitem[Hart \etal  1996]{machoref}
Hart, J. \etal  1996, { Publ. Astr. Soc. Pacific}, {\rm 108}, 220.

\bibitem[{Hudson} \& {Ebeling} 1997]{hudson+ebeling97}
{Hudson}, M.~J. \& {Ebeling}, H. 1997, { Astrophys. J.}, {\rm 479}, 621.

\bibitem[{Kim}, {Goobar}, \& {Perlmutter} 1996]{kimetal96}
{Kim}, A., {Goobar}, A., \& {Perlmutter}, S. 1996, { Publ. Astr. Soc. Pacific},
  {\rm 108}, 190.

\bibitem[Kolatt \& Bartelmann 1997]{kolatt+bartelmann97}
Kolatt, T.~S. \& Bartelmann, M. 1997, \mnras, {\it Submitted}.

\bibitem[Lauer \& Postman 1994]{lp94}
Lauer, T.~R. \& Postman, M. 1994, { Astrophys. J.}, {\rm 425}, 418.
\newblock [LP94].

\bibitem[Martin, Williams, \& Woodings 1997]{perth97}
Martin, R., Williams, A., \& Woodings, S. 1997, {\it IAU Circ \#6558}.

\bibitem[Marzke \etal  1995]{marzke95}
Marzke, R.~O., Geller, M.~J., da~Costa, L.~N., \& Huchra, J.~P. 1995, { Astron.
  J.}, {\rm 110}, 447.

\bibitem[Maza \etal  1997]{maza97}
Maza \etal  1997, {\it IAU Circ \#6531}.

\bibitem[Perlmutter \etal  1997]{perletal97}
Perlmutter, S. \etal  1997, { Astrophys. J.}, {\rm 483}, 595.

\bibitem[Phillips \& Davis 1995]{alinear95}
Phillips, A.~C. \& Davis, L.~E. 1995, in { Astronomical Data Analysis Software
  and Systems IV}, ed.\ R.A. Shaw, H.E. Payne, \& J.J.E. Hayes, volume~77 of {
  ASP Conference Series}, 297.

\bibitem[Phillips 1993]{phil93}
Phillips, M.~M. 1993, { Astrophys. J.}, {\rm 413}, L105.

\bibitem[Postman, Lauer, \& Strauss 1995]{pls}
Postman, M., Lauer, T.~R., \& Strauss, M.~A. 1995, {\it Private Communication}.
\newblock [PLS].

\bibitem[Reiss \etal  1997]{paperII}
Reiss, D., Germany, L., Schmidt, B.~P., \& Stubbs, C.~W. 1997, {\it In
  Preparation}.

\bibitem[Riess, Press, \& Kirshner 1995a]{rpk95b}
Riess, A., Press, B., \& Kirshner, R.~P. 1995a, { Astrophys. J.}, {\rm 438},
  L17.

\bibitem[Riess, Press, \& Kirshner 1995b]{rpk95}
Riess, A., Press, B., \& Kirshner, R.~P. 1995b, { Astrophys. J.}, {\rm 445},
  L91. \newblock [RPK95].

\bibitem[Riess, Press, \& Kirshner 1996]{rpk96}
Riess, A., Press, B., \& Kirshner, R.~P. 1996, { Astrophys. J.}, {\rm 473}, 88.

\bibitem[Riess \etal  1997]{riess+davis97}
Riess, A.~G., Davis, M., Baker, J., \& Kirshner, R.~P. 1997, \apj, {\it
  Submitted}.

\bibitem[Schechter, Mateo, \& Saha 1993]{dophotref}
Schechter, P.~L., Mateo, M., \& Saha, A. 1993, { Publ. Astr. Soc. Pacific},
  {\rm 105}, 1342.

\bibitem[Schmidt \etal  1997a]{schmidt97}
Schmidt, B.~P. \etal  1997a, \apj, {\it Submitted}.

\bibitem[Schmidt \etal  1997b]{highz97}
Schmidt, B.~P. \etal  1997b, { Bull. American Astron. Soc.}, {\rm 189(108)}.

\bibitem[{Strauss} \etal  1995]{strauss+etal95}
{Strauss}, M.~A., {Cen}, R., {Ostriker}, J.~P., {Lauer}, T.~R., \& {Postman},
  M. 1995, { Astrophys. J.}, {\rm 444}, 507.

\bibitem[Stubbs \etal  1993]{cameraref}
Stubbs, C.~W. \etal  1993, in { Proceedings of the SPIE}, volume 192, 1900.

\bibitem[Tammann 1994]{tammannrates94}
Tammann, G.~A. 1994, in { "Supernovae", Les Houches 1990, Session LIV}, ed.\
  S.~A. Bludman, R. Mochkovitch, \& J. Zinn-Justin, ), 3.

\bibitem[Treffers \etal  1997]{KAIT}
Treffers, R.~R., Peng, C.~Y., Filippenko, A.~V., \& Richmond, M.~W. 1997, {\it
  IAU Circ \#6627}.

\bibitem[Watkins \& Feldman 1995]{watkins+feldman95}
Watkins, R. \& Feldman, H.~A. 1995, { Astrophys. J.}, {\rm 453}, L73.

\end{thebibliography}

\newpage

\section{Figure Captions}
\label{sec:figcaptions}

\figcaption[reiss.fig1.ps]{
  The distribution of target clusters in the sky, in galactic (top)
  and equatorial (bottom) coordinates.  Filled square symbols
  represent clusters whose BCG distances were measured by LP94,
  filled circles were not. Thin circles enclose those clusters in
  which this project has found a supernova since June 1996 (several
  have two SNe).  Also shown are the direction of the LP94 bulk
  flow (LP), and the direction of the \protect\cite{CMBref}
  microwave background dipole (CMB).}
\label{fig:skydist}

\figcaption[reiss.fig2.ps]{
  The distribution in seeing FWHM on the 1.3m telescope at Mt.
  Stromlo over a year. Median seeing is $\sim 2.5^{\prime\prime}$.
  }
\label{fig:seeinghist}

\figcaption[reiss.fig3.ps]{
  A histogram of the number of days between each cluster observation
  ($\Delta t$). Top: The observations of June 1996 to the present;
  Bottom: A simulation of observations over a 3-year period.}
\label{fig:schedhist}

\figcaption[reiss.fig4a.eps,reiss.fig4b.eps,reiss.fig4c.eps,reiss.fig4d.eps,reiss.fig4e]{
  The search process illustrated on a subsection of the images in
  which SN 1996aj in Abell 3559 was detected. Fig. (a) shows a
  template, with a grid of 9 false sensitivity stars and masked
  saturated regions. The observation (b) is registered to the template
  (c). Fig. (d) shows the template of Fig. (a) after it has been
  convolved so that its PSF matches that of the registered image (c).
  After the registered image has been flux-matched to the convolved
  template, the template is subtracted. The residual is shown in Fig.
  (e), with the supernova (center) and five sensitivity stars clearly
  visible.}
\label{fig:searching}

\figcaption[reiss.fig5.ps]{
  A subraster of an SN candidate, produced by the search software.
  This is the actual subraster created for SN 1996aj seen in
  Fig.~\protect\ref{fig:searching}e. The supernova is clearly
  visible in both colors (SN), as are two cosmic rays (CRs,
  convolved during the subtraction process) in the V observation and
  subtraction.}
\label{fig:subraster}

\figcaption[reiss.fig6.ps]{
  Distribution of limiting magnitudes for all observations for which
  this was measurable. Typical median limits have been $\sim 20.4$
  in $R_{M}$ and $\sim 20.5$ in $V_{M}$.}
\label{fig:mlimhist}

\end{document}